\documentclass[useAMS,usenatbib]{mn2e}
 \usepackage{amsmath}
 \usepackage{deluxetable}
 \usepackage{graphicx} 
 \usepackage{upgreek}

\newcommand{\kepler}{{\em Kepler}}
\newcommand{\corot}{{\em CoRoT}}
\newcommand{\numax}{\mbox{$\nu_{\rm max}$}}
\newcommand{\Dnu}{\mbox{$\Delta \nu$}}

\newcommand{\muHz}{\mbox{${\upmu}$Hz}}

\newcommand{\half}{{\textstyle\frac{1}{2}}}

\def\note #1]{{\bf #1]}}

\title[Asymptotics of mixed modes in red giant stars]{Verification of asymptotic relation for mixed modes in red giant stars}

\author[C.~Jiang and J.~Christensen-Dalsgaard]{C.~Jiang$^{1}$\thanks{E-mail:jiangchen@phys.au.dk}
and J.~Christensen-Dalsgaard$^{1}$.\\
$^{1}$Stellar Astrophysics Centre, Department of Physics and Astronomy, Aarhus University, Ny Munkegade 120, DK-8000 Aarhus C, Denmark}

\begin{document}


\pagerange{\pageref{firstpage}--\pageref{lastpage}} \pubyear{2012}

\maketitle

\label{firstpage}

\begin{abstract}
High-precision space observations, such as made by the \kepler\ and \corot\ missions, allow us to detect mixed modes for $l = 1$ modes in their high signal-to-noise photometry data. By means of asteroseismology, the inner structure of red giant (RG) stars is revealed for the first time with the help of mixed modes. We analyse these mixed modes of a 1.3 ${\rm M}_{\sun}$ RG model theoretically from the approximate asymptotic descriptions of oscillations. While fitting observed frequencies with the eigenvalue condition for mixed modes, a good estimate of period spacing and coupling strength is also acquired for more evolved models. We show that the behaviour of the mode inertia in a given mode varies dramatically when the coupling is strong. An approximation of period spacings is also obtained from the asymptotic dispersion relation, which provides a good estimate of the coupling strength as well as period spacing when g-mode-like mixed modes are sufficiently dense. By comparing the theoretical coupling strength from the integral expression with the ones from fitting methods, we confirmed that the theoretical asymptotic equation is problematic in the evanescent region due to the potential singularities as well as the use of the Cowling approximation.
\end{abstract}


\section{Introduction}

As an excellent tool to study the structure of stars, asteroseismology has been developed rapidly in recent decades. By investigating stellar oscillations, it enables us to probe the interior  of stars. For solar-like stars, there are two main pulsation modes confined inside them: p modes and g modes. For the p modes, pressure is the restoring force and they are primarily vertical acoustic waves. For the g modes, buoyancy is the restoring force. p-mode oscillations are widely observed in main-sequence stars because of their large amplitudes and their motions are primarily vertical. However, although g-mode oscillations have much smaller amplitudes, which make them very hard to be observed directly, they are seen in terms of mixed modes in some more evolved stars. 

In main-sequence stars, p modes and g modes are trapped in their own cavities. When stars evolve to the giant stage, the very large gravitational acceleration in the core opens up the possibility that the two cavities could couple with each other, making these coupled modes have p-mode character in the outer part of stars and g-mode character in the core area, for which they are called \textit{mixed modes}. Thanks to the high quality of the photometry data from \kepler\ and \corot, many recent analyses of oscillations in red giant (RG) stars have been done \citep[e.g.][]{hek09,bed10,hub10,jia11,mat11,mos11,bau12,kal12}, which prove that using mixed modes is a robust tool to probe the inner structure and evolution of these evolved stars \citep{ bec11,bec12,bed11, mos12a}.

\cite{unn89} studied the general properties of stellar non-radial oscillations analytically using an asymptotic method and obtained eigenvalue conditions for p modes, g modes as well as mixed modes. \cite{jcd12} also analysed  the properties of the modes using an approximation to the asymptotic relations. In this paper, we introduce the typical properties of mixed modes and use models to predict the behaviour of these modes. 

\begin{figure}
\resizebox{1.0\hsize}{!}{\includegraphics[]{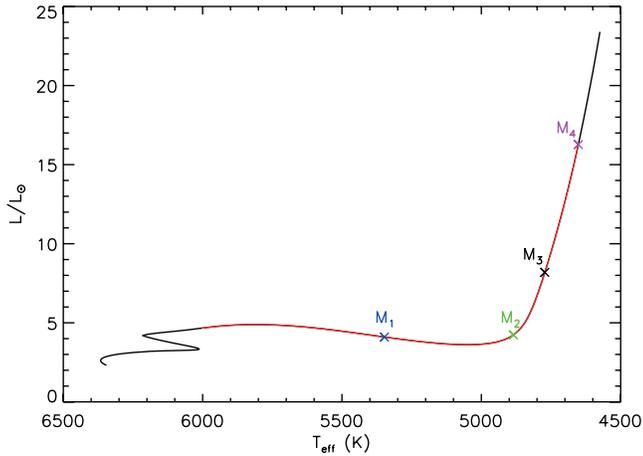}}
\caption{Evolution track of the $1.30 {\rm M}_{\sun}$ model. The red part covers all the modes shown in Fig.~\ref{fg:fevo}. The three crosses indicate the selected models coloured the same way as in Fig.~\ref{fg:fevo} plus an additional more evolved RG model $M_{\rm 4}$ (purple cross).}
\label{fg:track}
\end{figure}

\section{Asymptotic fitting of Mixed Modes}\label{sc:2.0}

When stars evolve to post-main-sequence stages, the large gravitational acceleration in the core increases the upper limit of the frequency range that g modes can reach, which results in the coupling of p modes and g modes. As a result, these coupled modes undergo so-called \textit{avoided crossing} and \textit{mode bumping}, their frequencies being shifted from the regular spacings indicated by the asymptotic descriptions. With the intention of asymptotic analysis, we follow a 1.3 ${\rm M}_{\sun}$ RG model with initial solar parameters that was calculated by the \textsc{ASTEC} evolution code \citep{jcd08a} and the \textsc{ADIPLS} oscillation package \citep{jcd08b}. Its evolutionary track is illustrated in Fig.~\ref{fg:track}. Several different cases of mixed modes will be discussed in the following sections.

\subsection{Asymptotic relations for p and g modes}\label{sc:2.1}

The frequencies of low-degree and high-order  p modes are regularly spaced, approximately following the asymptotic relation \citep{tas80,gou86}:
\begin{equation}
  \nu_{nl} = \frac{\omega_{nl}}{2{\uppi}}\simeq \Dnu (n + \half l + \epsilon_{\rm p}) - d_{nl}~,
        \label{eq:asyp}
\end{equation}
where $n$ is the radial order, $l$ is the angular degree of the mode and $\omega_{nl}$ is the angular frequency. $\Dnu$ is known as the large frequency separation, which is the inverse sound travel time across the star, given by
\begin{equation}
  \Dnu=\left(2\int_{0}^{R}\frac{{\rm d}r}{c}\right)^{-1}\; ,
      \label{eq:dnu}
\end{equation}
where $c$ is the sound speed and \textit{R} is the surface radius, and the integration is made over the distance $r$ to the centre. In equation~\eqref{eq:asyp}, $\epsilon_{\rm p}$ is a frequency-dependent phase shift due to the large gradient of the cut-off frequency near the stellar surface and $d_{nl}$ is a correction called the small separation. On the other hand, the periods of g modes satisfy an asymptotic relation \citep{tas80}:
\begin{equation}
  \Pi_{nl}=\frac{2{\uppi}}{\omega_{nl}} \simeq \Delta\Pi\left(n+\epsilon_{\rm g} \right)~,
  \label{eq:asyg}
\end{equation}
where $\epsilon_{\rm g}$ is again a phase shift, and 
\begin{equation}
  \Delta\Pi = \frac{2{\uppi}^2}{L}\left(\int_{r_1}^{r_2}N\frac{{\rm d}r}{r}\right)^{-1},
  \label{eq:dp}
\end{equation}
where $L=\sqrt{l(l+1)}$, $N$ is the Brunt-V\"ais\"al\"a frequency (or buoyancy frequency) and the integral is over the cavity where the g mode is trapped. However, the asymptotic relations for both p and g modes are only valid for high-order modes ($n\gg l$). 

\begin{figure}
\resizebox{1.0\hsize}{!}{\includegraphics[]{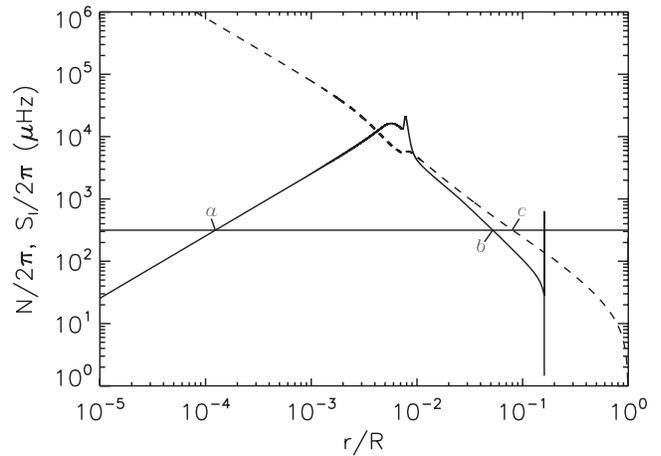}}
\caption{Propagation diagram of model $M_3$. The buoyancy frequency (solid curve) and Lamb frequency (dashed curve, $l = 1$) are shown in terms of corresponding cyclic frequencies, against fractional radius $r/R$. The axes are in logarithm. The horizontal line indicates the mode frequency $\nu = 315.26~\muHz$, which presents mixed-mode character. The locations of turning points of equation~\eqref{eq:k} are indicated by the letter $a$, $b$ and $c$, which divide the model into different propagation zones.}
\label{fg:propagation}
\end{figure}

The oscillation displacement $\boldsymbol{\delta r}$ can be separated into radial and horizontal components, $\xi_{\rm r}$ and $\xi_{\rm h}$,
\begin{equation}
\boldsymbol{\delta r} = \xi_{\rm r} \boldsymbol{a}_{\rm r} + \boldsymbol{\xi}_{\rm h}
\end{equation}
where $\boldsymbol{a}_r$ is a unit vector directed outward. \cite{unn89} introduced two variables $v$ and $w$ related to the displacement components as
\begin{equation}
v = \rho^{1/2}cr\left(\left|1-\frac{S_l^2}{\omega^2}\right|\right)^{-1/2}\xi_{\rm r}
\label{eq:v}
\end{equation} 
and
\begin{equation}
w=\rho^{1/2}r^2\omega^2\left(|N^2-\omega^2|\right)\xi_{\rm h},
\label{eq:w}
\end{equation}
with $\rho$ being the density. 
Neglecting the perturbation to the gravitational potential
\citep[the so-called Cowling approximation;][]{Cowlin1941},
the two variables $v$ and $w$ each approximately satisfy a second-order
differential equation that describes the behaviour of oscillations,
\begin{equation}
\frac{{\rm d}^2 v}{{\rm d} r^2}  + K^2 v = 0
\label{eq:difv}
\end{equation}
and
\begin{equation}
\frac{{\rm d}^2 w}{{\rm d} r^2} + K^2 w = 0
\label{eq:difw}
\end{equation}
with $K$ defined by
\begin{equation}
  K^2 \approx \frac{\omega^2}{c^2}\left(1-\frac{S_{l}^{2}}{\omega^{2}}\right)\left(1-\frac{N^{2}}{\omega^{2}}\right).
  \label{eq:k}
\end{equation}
Here $K^2$ is approximated by ignoring the acoustic cut-off frequency term that is generally small in the stellar interior and large near the surface. In equations ~\eqref{eq:v} and~\eqref{eq:k}, apart from $N$ there is another characteristic frequency $S_l$, also known as the Lamb frequency,
\begin{equation}
S_l^2=\frac{l(l+1)c^2}{r^2}.
\end{equation}
\cite{unn89} solved the set of differential equations ~\eqref{eq:difv} and \eqref{eq:difw} asymptotically to obtain $v$ and $w$ using the JWKB method (or Jeffreys, Wentzel, Kramers and Brillouin; \citealt{gou07}) and found that there is a general eigenvalue conditions for p and g modes ,
\begin{equation}
\int^{r_2}_{r_1}K {\rm d} r= {\uppi} ( n +\epsilon)\; .
\label{eq:phase}
\end{equation}
Here $n$ is the mode order, and $r_1$ and $r_2$ are adjacent turning points (where $K=0$), between which $K^2$ is positive. Again $\epsilon$ is a phase correction that depends on the structure of the model near the turning points.  The dependence of $\epsilon$ on frequencies is implied in the definition of $K$, which can be explored numerically by using equation~\eqref{eq:phase} \citep{jcd84}. 
We note that the derivations of equations (\ref{eq:difv}) and (\ref{eq:difw}) 
neglected terms that may lead to singularities at critical points in the model,
although typically in {\it evanescent} regions where $K^2 < 0$.
This is explicit in a second-order equation,
exact in the Cowling approximation, presented by \citet{Gough1993},
which is of the same form as equations (\ref{eq:difv}) and (\ref{eq:difw})
but expressed in terms of a different dependent variable.

\begin{figure}
\resizebox{1.0\hsize}{!}{\includegraphics{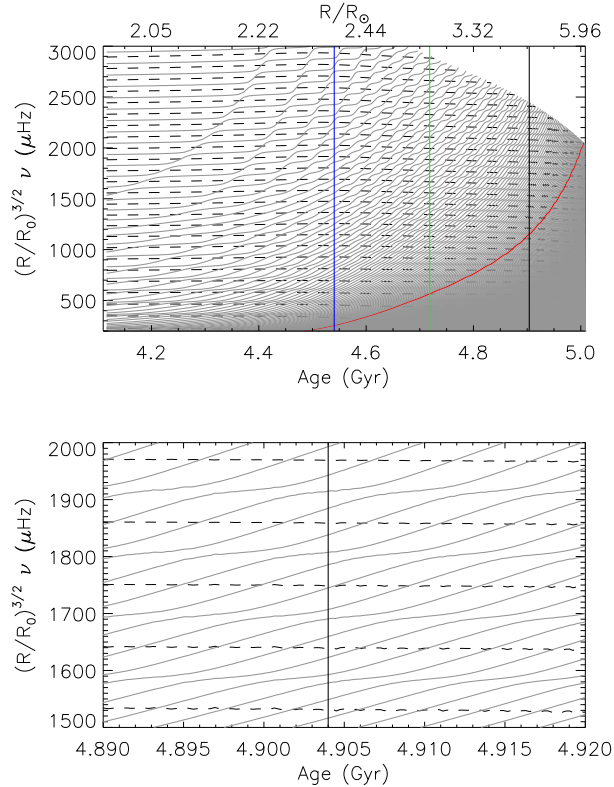}}
\caption{Scaled oscillation frequencies as a function of age and radius in the 1.30 ${\rm M}_{\sun}$ model. The models in the upper panel start from the subgiant to the RG stage, while the lower panel shows a small segment of the solutions around the age of 4.904 Gyr. On the $y$-axis, $R$ is the surface radius of the model and $R_0$ is the zero-age main-sequence radius. Modes with the same radial order have been connected in the same line. The dashed lines are for radial modes and the solid lines for $l = 1$. The three vertical lines indicate three models. Together with the last model they are discussed in detail, and their locations are shown in the evolution track of Fig.~\ref{fg:track}. Evolution of one g mode with $n=-50$ is highlighted as a red curve and is described in Section~\ref{sec:class}. The computations of frequencies stop at the acoustic cut-off frequency, which leaves the blank area in the upper figure.}
\label{fg:fevo}
\end{figure}

\begin{table}
\caption{Fundamental parameters of the 1.30 ${\rm M}_{\sun}$ giant models.
$R$ and $L$ are surface radius and luminosity, provided in units of the solar
values, $T_{\rm eff}$ is effective temperature,
$\nu_{\rm max}$ is the estimated frequency at maximum power and $n_{\rm max}$
is the corresponding radial-mode order, $\Delta \nu$ is the acoustic-mode
frequency spacing obtained from fitting the frequencies and 
$\Delta \nu_{\rm int}$ was determined from the asymptotic integral,
equation~(\ref{eq:dnu}), and $\Delta {\uppi}_{\rm int}$ is the asymptotic period spacing
of dipolar modes (cf. equation~\ref{eq:dp}).}
\label{tb:models}
\centering
\begin{tabular}{@{}lcccc}
\hline
\small{${\rm Model}$} & \small{$M_{\rm 1}$} & \small{$M_{\rm 2}$} & \small{$M_{\rm 3}$}  &  \small{$M_{\rm 4}$} \\
\hline
\small{$R/R_{\sun}$} & \small2.36 & \small2.88 & \small4.20 & \small6.22\\
\small{$T_{\rm eff}$ (K)} & \small5347.8 & \small4884.9 & \small4772.2  & \small4651.9\\
\small{$L/L_{\sun}$} & \small4.101 & \small4.239 & \small8.189  & \small16.272\\
\small{$\numax$ (\muHz)} & \small745.24  & \small513.78 & \small252.17  & \small117.62\\
\small{\mbox{$n_{\rm max}$}} & \small16 & \small15 & \small13  & \small11\\
\small{$\Dnu$ (\muHz)}  & \small43.38 & \small31.62 & \small17.88  & \small9.89\\
\small{$\Dnu_{\rm int}$ (\muHz)}  & \small45.32 & \small33.19 & \small18.79  & \small10.38\\
\small{$\Delta{\rm {\uppi}}_{\rm int}$ (s)} & \small191.69 & \small111.73 & \small 87.78 & \small75.50\\
\hline
\end{tabular}
\end{table}

\subsection{Asymptotic relation for mixed modes}\label{sc:2.2}

\begin{figure*}
\resizebox{1.0\hsize}{!}{\includegraphics{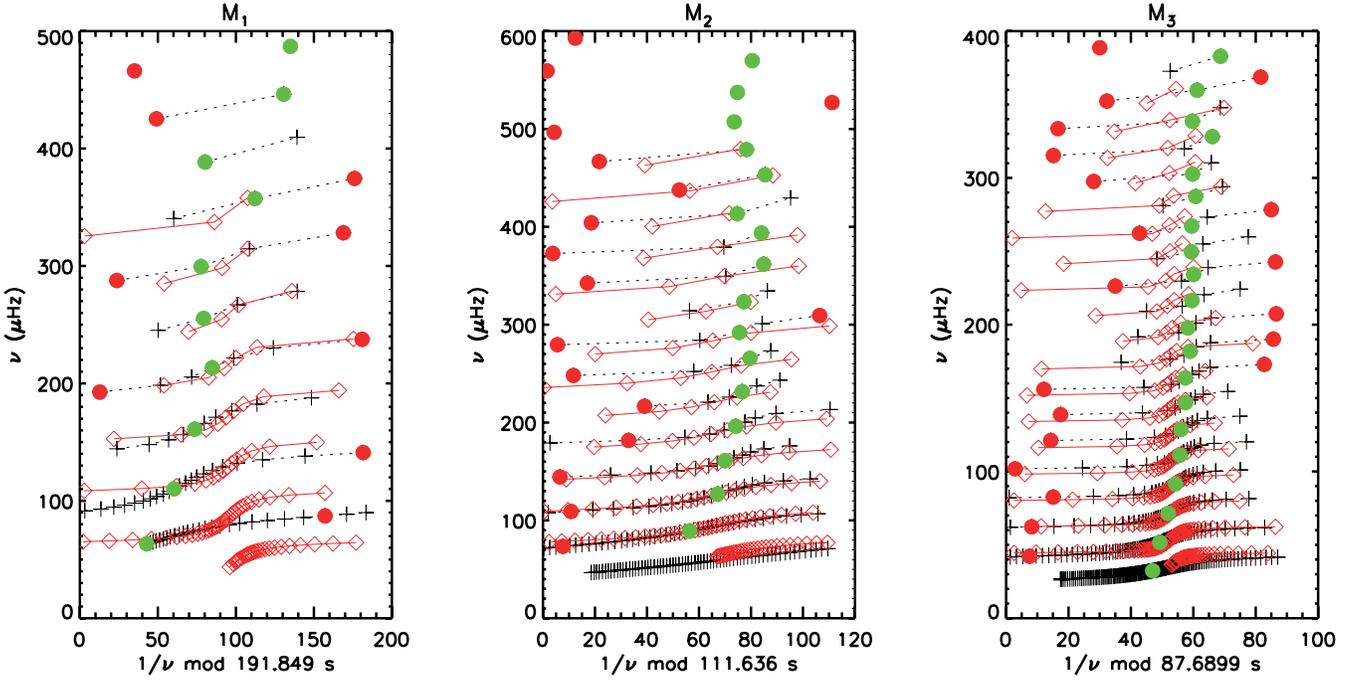}}
\caption{Period \'{e}chelle diagrams for the three models $\rm M_1 \ - \ \rm M_3$. The $x$-axis presents the period modulo the period spacing $\Delta{\uppi}$ gained from fitting and the $y$-axis is the frequency. Crosses linked by dashed lines indicate the theoretical frequencies. Diamonds correspond to the asymptotic fit. Green circles are the most g-m-like mixed modes while red circles indicate the most p-m-like mixed modes.}
\label{fg:fitting}
\end{figure*}

In order to discuss a similar asymptotic relation for mixed modes, the propagation diagram of model $M_3$ (see Section~\ref{sc:fmm} for details of the model) is shown in Fig.~\ref{fg:propagation} with eigenfrequency of 315.26 $\muHz$ indicated by the horizontal line that intersects with the characteristic frequencies at turning points `$a$', `$b$' and `$c$'. In this case, g modes are trapped in the interval between the layer very close to the innermost point `$a$' and `$b$', while p modes travel in the outer part of the star between `$c$' and the surface, where $K^2$ is positive for both cases. The region between `$b$' and `$c$' is called the evanescent region where $K^2$ is negative and the eigenfunctions behave exponentially. Equation~\eqref{eq:phase} can be considered as an eigenvalue condition for p and g modes \citep{unn89}. For g modes:

\begin{equation}
\int^{r_{\rm b}}_{r_{\rm a}}K{\rm d}r \approx {\uppi}(n+1/2+\epsilon_{\rm g}),
\label{eq:gcon}
\end{equation}
and for p modes:
\begin{equation}
\int^{R}_{r_{\rm c}}K{\rm d}r \approx {\uppi}(m+\epsilon_{\rm p}),
\label{eq:pcon}
\end{equation}
where $n$ and $m$ are integers that also define the orders of p and g modes, individually. In the present case, with a radiative core, $r_{\rm a}$ is very close to the centre. If there is a convective core, the deeper boundary of g modes is outside the core. 

\cite{unn89} describe an eigenvalue condition for mixed modes using
a coefficient $q$ measuring the coupling strength between gravity-wave and acoustic-wave cavities:
\begin{equation}
\cot\left(\int^{r_{\rm b}}_{r_{\rm a}}K{\rm d}r\right)\tan\left(\int^R_{r_{\rm c}}K{\rm d}r\right)= q \; .
\label{eq:mix}
\end{equation}
[A qualitatively similar relation was obtained by \citet{jcd12} from the 
analysis of a simple toy model.]
As discussed below, equation (\ref{eq:mix}) has been a very powerful tool in 
the analysis of mixed modes in RGs \citep[e.g.,][]{mos12a},
with the coupling strength $q$ obtained by fitting the expression to 
observed or computed frequencies of oscillation.

It is of obvious interest to investigate how the coupling strength reflects
the properties of the stellar interior.
From a simple analysis of equations (\ref{eq:difv}) -- (\ref{eq:k}), \citet{unn89}
obtained the estimate
\begin{equation}
q_{\rm int}=\frac{1}{4}\exp\left(-2\int^{r_{\rm c}}_{r_{\rm b}}|K|{\rm d}r\right) \;.
\label{eq:qori}
\end{equation}
We note that this is obviously questionable, given the problems with 
singularities discussed above.
Even so, we find it interesting in the following to compare $q_{\rm int}$ as
computed from equation (\ref{eq:qori}) with the results of frequency fits.

When the coupling is very weak, $q$ is close to $0$. It is reasonable to estimate that $q$ is a small quantity and there is a condition that satisfies equation~\eqref{eq:mix}:
\begin{equation}
\left. \begin{aligned}
  \int^{r_{\rm b}}_{r_{\rm a}}K{\rm d}r \approx {\uppi}(n+1/2 \pm \epsilon)\\
  \int^{R}_{r_{\rm c}} K{\rm d}r \approx {\uppi}(m \mp \epsilon)
  \phantom{\hspace{1cm}}
\end{aligned}
  \right\}.
  \label{eq:mixcon}
\end{equation}
It should be noted that equation~\eqref{eq:mix} is the general condition even for pure p and g modes if we take $q = 0$. When mixed modes occur, the small value of $q$ leads to deviations from pure oscillation modes, which are indicated by corresponding $\pm$ and $\mp$ signs in equation~\eqref{eq:mixcon}. From equation~\eqref{eq:k} $K$ only depends on the frequency for a given model. So the occurrence of oscillation modes is also determined by the frequency. Hence let us consider a model with frequency gradually increasing. In the very beginning, the frequencies are very small making the integrals in equation~\eqref{eq:gcon} so great that a large number of g modes satisfy equation~\eqref{eq:gcon}. However, there is no p mode existing at the very low frequency range because $S_l \gg \omega^2$.  As the frequency increases, the region in which gravity waves are trapped becomes narrower and therefore the integral of $K$ monotonically decreases. But once the frequency is greater than $S_l$ the integral over acoustic-wave zone monotonically increases. If the frequency is high enough to make gravity and acoustic waves couple with each other meaning that the integrals of $K$ approach $(n-1/2){\uppi}$ and $(m+1){\uppi}$ at the same time, the eigenvalue condition equation~\eqref{eq:mixcon} is satisfied and an avoided crossing occurs. When the frequency increases beyond $N$, only equation~\eqref{eq:pcon} can be satisfied by p modes.

Fig.~\ref{fg:fevo} shows the evolution of the oscillation modes for the 1.30 ${\rm M}_{\sun}$ model, as functions of stellar age and radius. The frequencies in the figure have been scaled according to the inverse of dynamical time-scale $t^{-1}=(R^3/GM)^{-1/2}$, which makes the frequencies of acoustic modes vary little with time. There are clear uniform spacings between  p modes. On the other hand, the scaled frequencies of gravity modes (see solid lines at the lower-left corner)  have an increasing trend with age, which is a consequence of the increase in $N$. However, the striking feature of the frequencies in Fig.~\ref{fg:fevo} is the interaction between the dipolar acoustic and gravity modes. At an early age when the star just leaves the main sequence, the high-order modes are only pure p modes, while g modes are located at low frequency but gradually increasing with time. When the frequencies of the g modes are high enough to interact with p modes, the horizontal lines of p modes are bumped up to relatively higher values, which breaks the equally spaced pattern of p modes. These interactions take place through a series of  avoided crossings when the two modes exchange nature. On the other hand, as the star ages, g modes dominate the whole frequency range, while p modes can only be revealed in terms of mixed modes. In this case, frequencies will be diminished when avoided crossings happen, which form the very dense p-dominated mixed modes (see Section~\ref{sec:class}) in the time sequence that still can be observed as very dense horizontal lines in Fig.~\ref{fg:fevo}. 

\section{Classification of mixed modes}
\label{sec:class}

\subsection{Fitting mixed modes}\label{sc:fmm}
While a mode is undergoing avoided crossing its frequency deviates a little from its original value which makes $K$ satisfy condition equation~\eqref{eq:mixcon}. One can calculate $K$ approximately in different regions separated by turning points ($a$, $b$ and $c$ in Fig.~\ref{fg:propagation}) \citep[see][]{jcd12}.It should be noted that these approximations are valid except near the turning points. Therefore the integral of $K$ in the outer region is closely related to $\Dnu$  and that in the inner region  to $\Delta\Pi$  in this asymptotic analysis (see equations~\ref{eq:dnu} and equation~\ref{eq:dp}). Substituting $\Delta\Pi$ and $\Dnu$ into equation~\eqref{eq:mix} yields equation (9) in \cite{mos12a}:
 \begin{equation}
 \nu = \nu_{n_{\rm p},\,l=1} + \frac{\Dnu}{{\uppi}}\arctan \left[q\tan {\uppi} \left(\frac{1}{\Delta \Pi_1 \nu}-\epsilon \right) \right],
 \label{eq:mixnu}
 \end{equation}
where $\nu_{n_{\rm p},\,l=1}$ is the uncoupled solutions of p modes and $\epsilon$ is the phase shift in equation~\eqref{eq:asyg} which makes the obtained periods close to $(n+1/2+\epsilon)\Delta\Pi_l$ when the coupling is weak \citep{mos12a}. Equation~\eqref{eq:mixnu} is the expression for the $l=1$ mixed modes coupled to the pure p modes $\nu_{n_{\rm p},\,l=1}$. It gives a more intuitive view of the frequency change from pure p modes with several observable seismic parameters than equation~\eqref{eq:mix} that demands the knowledge of the structure of stars. However, a similar expression for p-dominated mixed modes coupled with pure g modes can also be acquired, though the asymptotic relations break down for low radial order modes. But we only focus on the former case in this paper. With the assumption of the asymptotic relation for p modes, one can approximate  $\nu_{n_{\rm p},\,l=1}$ which can be affected by the correction term $\epsilon$ to some extent. $\Dnu$ is obtained as the mean value of frequency spacings of radial modes, assumed also to be valid for dipole modes. The frequencies $\nu$ are either from models or observations. 

\begin{figure}
\centering
\resizebox{0.6\hsize}{!}{\includegraphics{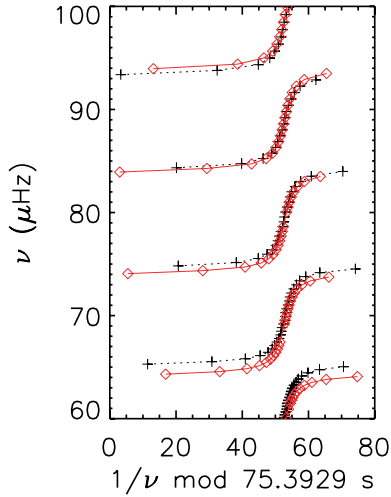}}
\caption{Enlarged period \'{e}chelle diagram for model $M_{\rm 4}$.}
\label{fg:m4fit}
\end{figure}

Here, we only analyse theoretical modes and use those around the frequency $\numax$ of maximum power, namely within the range $[\numax - 3\Dnu, \numax + 3\Dnu]$. Since $\numax$ is not given directly from models, it is scaled from the solar value using the usual scaling relations \citep{hans95}. Hence, $\Delta\Pi$, $q$ and $\epsilon$ are left as open parameters and their values are gained after a least-squares fit to frequencies by the method of grid searching. We constructed grids of $q$ and $\epsilon$ from 0 to 1 with a step of 0.01 and 0.1, respectively. $\Delta\Pi$ is searched around a preliminary value, which is estimated by equation~\eqref{eq:dp} when fitting only theoretical frequencies, within a small range ($[\Delta\Pi - 3 \; \rm s, \Delta\Pi + 3 \; \rm s]$). However, if we are confronting observed frequencies, $\Delta\Pi$ is estimated between $\Delta\Pi_{\rm obs}$ and $3\Delta\Pi_{\rm obs}$, where $\Delta\Pi_{\rm obs}$ is the measured value of the period spacing between bumped mixed modes and significantly smaller than $\Delta\Pi$. In this case, more computing time is required, but this is beyond the scope of this paper. In reality, the coupling strength varies with frequency, but the result from this asymptotic fitting is the mean value $\langle q \rangle$.

\begin{figure}
\resizebox{1.0\hsize}{!}{\includegraphics{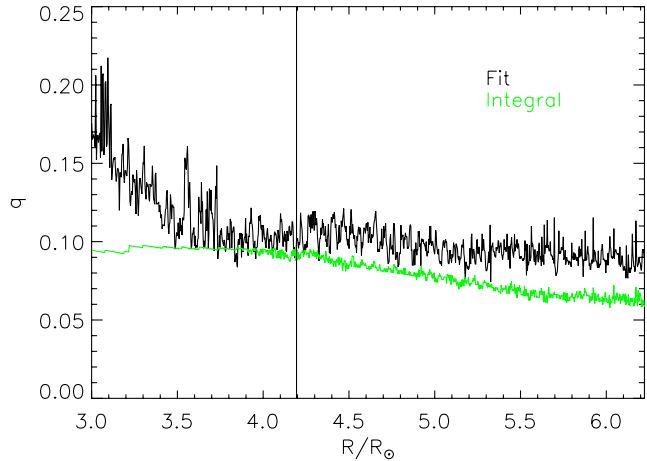}}
\caption{Coupling strength obtained from different means indicated by different coloured lines against radius covering all the ascending-branch models in Fig.~\ref{fg:fevo}. The `Fit' stands for the asymptotic fitting method (the same as in Table~\ref{tb:fitting}). The `Integral' results are calculated from the approximate integral in equation~\eqref{eq:qori} and they are the average value of  $\nu_{n_{\rm p},\,l=1}$ modes (see Table~\ref{tb:dpfit}). Only modes over the interval $[\numax - 3\Dnu, \numax + 3\Dnu]$ are taken into account for a given model for both methods. The vertical line indicates the location of model $M_{\rm 3}$, and $M_{\rm 4}$ is at the end of the figure.}
\label{fg:q}
\end{figure}

We selected four models at different locations on the evolution track (see Fig.~\ref{fg:track}). The first model ($M_{\rm 1}$) is in the middle of the subgiant branch when p modes are still dominating the high-frequency range. The second model ($M_{\rm 2}$) lies at the base of RG branch as low-order g modes penetrate into the p-mode area in frequency. The third model ($M_{\rm 3}$) is on the ascending branch so that g modes are reigning over the frequency range with p modes penetrating in them. The last one ($M_{\rm 4}$) is farther up on the ascending branch. Their fundamental parameters are given in Table~\ref{tb:models}. For the seismic parameters, the large frequency separations $\Dnu$ are the mean value taken from radial modes, the period spacings of dipole g modes $\Delta\Pi_{\rm int}$ are the theoretical values derived from the integral of $N$ (equation~\ref{eq:dp}) while similarly $\Dnu_{\rm int}$ are from the integral of the inverse of the sound speed (equation~\ref{eq:dnu}), and $n_{\rm max}$ is the radial order at $\numax$. Since g modes are equally separated in period, it is reasonable to plot them in a period \'{e}chelle diagram, which is just like the classical \'{e}chelle diagram for p modes, but the $x$-axis is defined as the period modulo the period spacing $\Delta\Pi$. Fig.~\ref{fg:fitting} shows asymptotic fits to theoretical frequencies of the first three models. Those g-mode-like modes are located in the middle of the pattern, at $1/\nu=\Delta\Pi/2$ (modulo $\Delta\Pi$) if $\epsilon$ is $0$. As noted by \cite{bed11}, the S-pattern per $\Dnu$-wide interval observed in Fig.~\ref{fg:fitting} is the outcome of coupling. A higher coupling strength would result in gentle central patterns in each segment. The qualitative agreement of the fit to $M_{\rm 1}$ mixed modes is not good except for medium radial orders, as expected. This is because $q$ varies with frequency and the mean value $\langle q \rangle$ used in the fitting usually deviates from the real one of each segment. The agreement gets better as the model evolves (see the additional blown-up example of $M_{\rm 4}$ in Fig.~\ref{fg:m4fit}), but the fitting gives an abnormally high $q$ for $M_{\rm 1}$ and its vicinity (Table~\ref{tb:fitting}). Hence, for subgiant models, our obtained mean value of $q$ is not close to the actual coupling coefficient for low-radial-order g mixed modes, which is the reason that this asymptotic fitting method is not suitable for mixed modes in subgiant models. \cite{ben12} introduced a way to fit mixed modes in subgiant stars\footnote{They found that the coupling strength of the dipole mixed modes is predominantly a function of stellar mass and appears to be independent of metallicity.}. However, the fitting provides reasonable $q$ and $\Delta\Pi$ for models on the ascending branch. 

\begin{figure}
\resizebox{1.0\hsize}{!}{\includegraphics{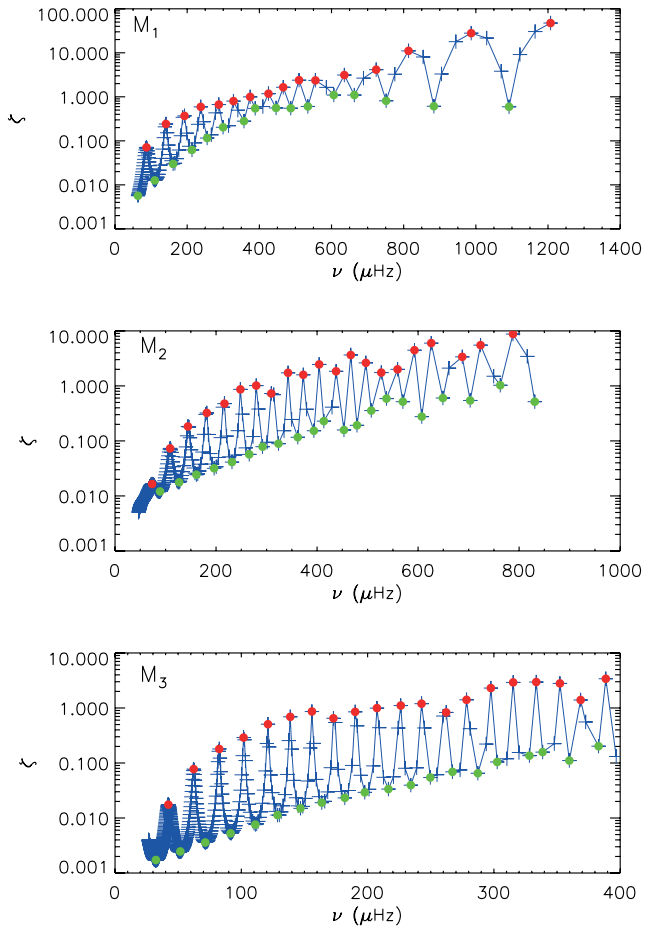}}
\caption{The ratio of dimensionless mode inertia in the p cavity over the one in the g cavity of the three models, as a function of frequency. The blue crosses represent the ratios for each oscillation mode. Green and red circles correspond to the same kinds of mixed modes as in Fig.~\ref{fg:fitting}.}
\label{fg:zeta}
\end{figure}

Fig.~\ref{fg:q} collects the results along the evolution sequence, 
comparing averages of the fitted values of $q$ with the approximate value
obtained from the integral in equation (\ref{eq:qori}). 
As indicated, the fitting results for $q$ are somewhat larger than the values 
of $q_{\rm int}$ in more evolved models.
This is hardly surprising, given the problems associated with
the simplified asymptotics, particularly as used in the evanescent region.
Concerning these obvious limitations, a relatively simple test is to 
redo the analysis for frequencies computed in the Cowling approximation.
We have done so for model $M_4$, at the right-hand edge of Fig. \ref{fg:q}.
The average $q$ resulting from the fits to frequencies computed in the Cowling
approximation is lower by about 0.02 than when using frequencies computed 
with the full equations;
this, probably coincidentally, brings the results of the fit substantially 
closer to $q_{\rm int}$.
A more detailed analysis is beyond the scope of this paper, however.
A solution of $\epsilon$ can also be found after the fitting. Although it decides the absolute position of the pattern in the period \'{e}chelle diagram, it 
has little influence on the solutions of the fitting \citep{mos12a}.    

\begin{table}
\caption{Results,
averaged over the interval $[\numax - 3\Dnu, \numax + 3\Dnu]$,
of fitting the asymptotic relation, equation (\ref{eq:mixnu}), to model frequencies.
}
 \label{tb:fitting} 
 \centering
\begin{tabular}{@{}lcccc}
\hline
\small{Model} & \small{$M_{\rm 1}$} & \small{$M_{\rm 2}$} & \small{$M_{\rm 3}$} & \small{$M_{\rm 4}$}\\
\hline
\small{$\Delta\Pi$ (s)} & \small191.74 & \small111.98 & \small 88.80 & \small 75.58\\
\small{$q$} & \small0.40 & \small0.23 & \small0.12 & \small 0.10\\
\small{$\epsilon$} &\small0.8 &\small0.1 & \small 0.2  & \small 0.27\\
\hline
\end{tabular}
\end{table}

From the computed models, we can obtain the gravity-mode period spacing $\Delta\Pi$ as well as the coupling strength $q$ through the fitting to theoretical mixed modes. Furthermore, with the help of models we can study properties of mixed modes by observing some parameters that characterize oscillation modes, such as kinetic energy \citep{unn89} or mode inertia \citep{dzi01}. We computed the normalized mode inertia given by \citep{jcd08b}:
\begin{equation}
E=\frac{\int_0^R\left(\xi_{\rm r}^2+L^2\xi_{\rm h}^2\right)\rho r^2{\rm d} r}{M\xi_{\rm r}(R)^2},
\end{equation} 
where $\rho$ is the density. For a mode that is dominated by p-mode nature, the inertia is contributed mostly by the p cavity, which is mainly from $\xi_{\rm r}$. For a mixed mode that behaves predominantly g-mode like, the inertia would be very large owing to the high density in the core area and hence is dominated by $\xi_{\rm h}$ in the g cavity \citep{bookas}.  Therefore it is convenient to measure the nature of mixed modes with the ratio of mode inertia in the p cavity over that in the g cavity:

\begin{equation}
\zeta=\frac{E_{\rm p}}{E_{\rm g}} \approx \frac{\int_{r_{\rm c}}^R \xi_{\rm r}^2 \rho r^2{\rm d} r}{\int_{r_{\rm a}}^{r_{\rm b}} L^2\xi_{\rm h}^2\rho r^2{\rm d} r}.
\end{equation}

\begin{figure}
\resizebox{1.0\hsize}{!}{\includegraphics{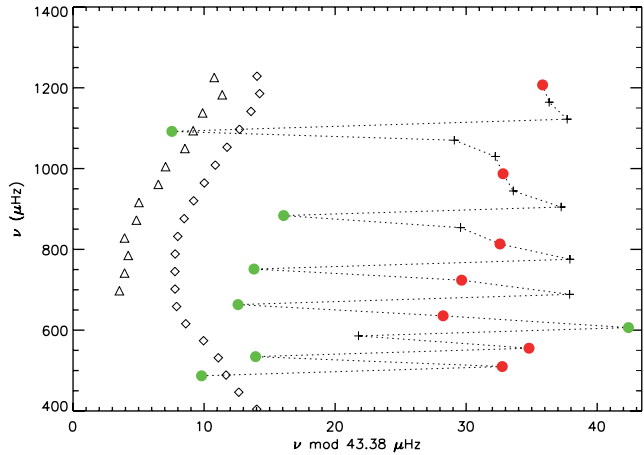}}
\caption{Frequency \'{e}chelle diagram of oscillation modes ($n > 0$) in $M_{\rm 1}$. Modes of different degrees are indicated by diamonds ($l = 0$), crosses ($l = 1$) and triangles ($l = 2$). Green and red circles correspond to the same kinds of mixed modes as in Fig.~\ref{fg:fitting}. To guide the eyes, the dotted line connects the dipolar mixed modes. Mixed modes for $l = 2$ modes are suppressed in the plot.}
\label{fg:m1p}
\end{figure}

Unlike \cite{mos12b} dividing mixed modes into two categories by the size of the amplitude in the core (see also \citealt{goup13}), we divide them by the local extreme values of $\zeta$, namely those mixed modes having the local maxima of $\zeta$ and their vicinities are called p-m modes and those at the local minima correspond to g-m modes. Although low-frequency mixed modes are principally dominated by g-mode nature ($\zeta \sim 0$), they are always coupling with p modes at different levels. When the coupling effect is strong, the mode tends to have more p-mode nature which leads to an increase in $\zeta$ but decrease in frequency. On the other hand, for larger frequency modes where the p character dominates and $\zeta$ is very large ($\zeta \gg 1$), their frequencies would be bumped up to a higher value but their $\zeta$ values are diminished when the coupling with the g-mode character is strong. The frequencies of these p-m modes are close to the frequencies of pure dipolar p modes, because they are little affected by the g cavity, and they can be observed clearly. Similarly, g-m modes are close to pure g modes. Additionally, mixed modes that have nearly equal mode inertia contributions from the envelope and the core take values of $\zeta$ of the order of $1$. They also have a significant g component that is valuable for stellar interior research, and may also be observed directly. The variation of $\zeta$ in dipolar modes with frequency is shown in Fig.~\ref{fg:zeta}, p-m and g-m modes being illustrated by coloured dots.

Since $M_{\rm 1}$ is a subgiant model, there exist high-order p modes as well as g modes. For mixed modes in the g-mode frequency range, g-m modes are expected to align vertically in the middle of the S-pattern in the period \'{e}chelle diagram (green circles in Fig.~\ref{fg:fitting}), while modes affected by p-mode character shift sidewards as p-m modes locating at the edge of the S-pattern (red circles in Fig.~\ref{fg:fitting}). For higher frequency mixed modes which have significant p-mode nature,  p-m modes line up more or less vertically as radial modes do in frequency \'{e}chelle diagram (red circles in Fig.~\ref{fg:m1p}) while modes affected by g-mode character shift sidewards as g-m modes positioning at the edge of each pattern (green circles in Fig.~\ref{fg:m1p}). The reason for the g-m modes in $M_{\rm 1}$ not lining up well is that the radial orders of these modes are relatively low and hence the asymptotic theory is not quite valid. As seen in $M_{\rm 3}$ of Fig.~\ref{fg:fitting}, the alignment is much better as there are adequate high-radial-order modes. 

\begin{figure}
\resizebox{1.0\hsize}{!}{\includegraphics{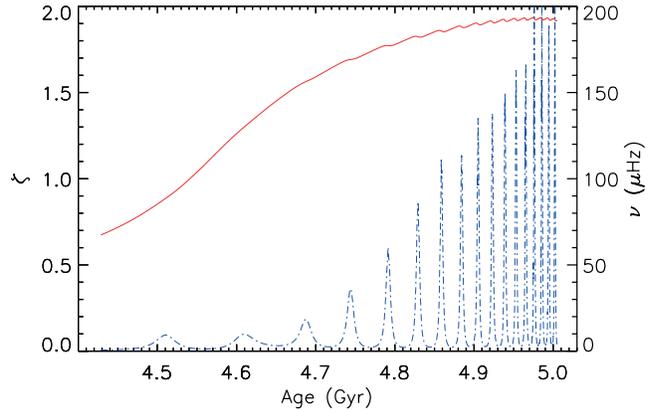}}
\caption{The evolutionary view of frequency (red solid line) and $\zeta$ (blue dash-dotted line) of a g mode ($n = -50$). The frequencies are indicated by the right $y$-axis.}
\label{fg:50}
\end{figure}

In summary, when the coupling is strong, g-dominated mixed modes usually have small values of $\zeta$ which increase because of the kinetic energy contributions from the envelope increase. In contrast, the large $\zeta$ values of p-dominated mixed modes decrease as a result of the effect by g-m modes. This is clearly seen by the study of the evolutionary variations of a g mode ($n = -50$), in terms of frequency and $\zeta$ (Fig.~\ref{fg:50}). The frequency of the mode increases smoothly at an early age, though there are some tiny influences from p modes which make $\zeta$ grow a little but remain small ($\zeta \ll 1$). When it ages to around 4.8 Gyr, the frequency decreases at the point where the local maximum of $\zeta$ is close to 1, which means the effect of avoided crossing is much greater than earlier. And a series of crossings followed with decreasing period means the p-mode character is becoming obvious because of the increasing coupling, and the possibility of this former pure g mode being detected is increasing too.

\section{Period spacing}

\begin{figure}
\resizebox{0.9\hsize}{!}{\includegraphics{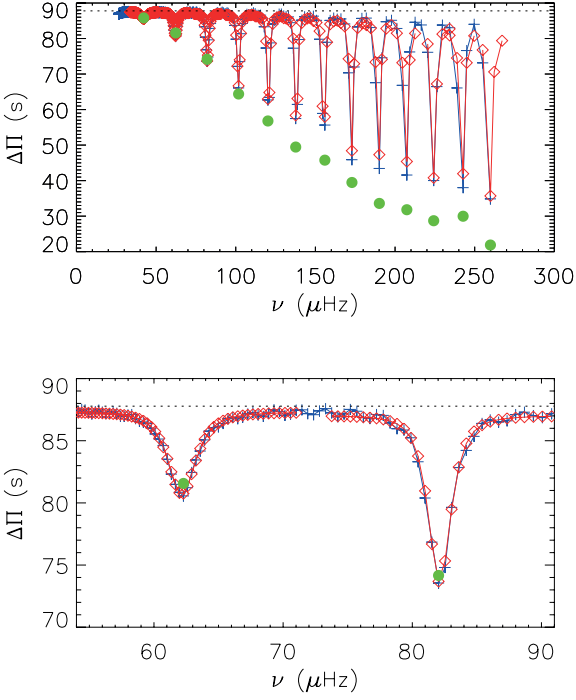}}
\caption{Period spacings for g modes of model $M_{\rm 3}$ (blue crosses) against angular frequency, with solutions to the asymptotic expression equation~\eqref{eq:dpfit} (red diamonds). Green circles are minimum spacings in each segment computed from equation~\eqref{eq:mindp} using $q_{\rm fit}$ listed in Table~\ref{tb:dpfit}. The horizontal dotted line is the theoretical period spacing of pure g modes defined by equation~\eqref{eq:dp}. The lower panel shows detailed properties of the second and third segments.}
\label{fg:dpfit}
\end{figure}

\begin{table}
\caption{Oscillation parameters $\omega_{\rm p}, \omega_{\rm g}$
and $q$ for model $M_{\rm 3}$
obtained by fitting computed period spacings to equation (\ref{eq:dpfit}),
for individual acoustic resonances characterized by $\nu_{\rm p\hbox{-}m}$.
$\Delta \Pi$ and $\Dnu$ were calculated from the individual values of
$\omega_{\rm g}$ and $\omega_{\rm p}$.
For comparison, we also show $q_{\rm int}$ calculated from equation (\ref{eq:qori})
at each $\nu_{\rm p\hbox{-}m}$, and $q_{\rm fit}$ resulting from fitting frequencies within
[$\nu_{\rm p\hbox{-}m} - \frac{1}{2}\Dnu$, $\nu_{\rm p\hbox{-}m} + \frac{1}{2}\Dnu$]
to equation (\ref{eq:mixnu}).}
\label{tb:dpfit}
\begin{tabular}{@{}cccccccc}
\hline
\small{$\nu_{\rm p\hbox{-}m}$} &  \small{$\omega_{\rm g}$} & \small{$\omega_{\rm p}$} &  \small{$\Delta \Pi$} &  \small{$\Dnu$} &  \small{$q$}  &  \small{$q_{\rm int}$} &  \small{$q_{\rm fit}$}  \\
 \small{($\muHz$)} &  \small{($\rm s^{-1}$)} & \small{($\muHz$)} &  \small{(s)} &  \small{($\muHz$)} & & &  \\
\hline
\small 42.22  & \small0.226    & \small42.10    & \small 87.51      & \small 21.05     & \small0.33         & \small 0.031    & \small0.38\\
\small 62.28     & \small0.226       & \small41.42      & \small 87.30       & \small 20.71    & \small0.20         & \small 0.053  & \small0.25\\
\small 82.04    &  \small0.227     & \small41.05    &  \small 87.02     &  \small 20.53    &    \small0.16     &   \small  0.070  &  \small0.18\\
\small 101.79   & \small0.228    &  \small40.67    &  \small 86.72     &  \small 20.34    &  \small0.15     &  \small 0.084   &  \small0.14\\
\small 120.19    &  \small0.229     &  \small40.20    &  \small 86.32     &  \small 20.10    &  \small0.14     &  \small 0.096   & \small0.13\\
\small137.64      &  \small0.230     &  \small39.44    &  \small 85.87    &  \small 19.72     &  \small0.14     &  \small 0.107   &  \small0.12\\
\small155.92     &  \small0.232     &  \small38.82     &  \small 85.19    &  \small 19.41     &  \small0.14      &  \small 0.116    &  \small0.13\\
\small172.93     &  \small0.233      &  \small 38.36    & \small 84.77     &  \small 19.18    & \small0.16      &  \small 0.123   &  \small0.12\\
\small190.14     &  \small0.235     &  \small 37.93     &  \small 84.09    &  \small 18.97     &  \small0.17       &  \small 0.129   &  \small0.11\\ 
\small207.39    &  \small0.240    &  \small37.60    &  \small 82.36     &  \small 18.80   &  \small0.17    &  \small 0.134   &\small0.12\\
\small224.24     &  \small0.238     &  \small37.37    &  \small  83.06    &  \small 18.68   &  \small0.22    &  \small 0.139  &  \small0.12\\  
\small242.71    &  \small0.242    &  \small37.23     &  \small 81.49     &  \small 18.62     &  \small0.20      &  \small 0.144   &  \small0.15\\
\small259.86     &  \small0.246     &  \small37.05    &  \small 80.12    &  \small 18.53     &  \small0.20    &  \small0.149    &  \small0.11\\ 
\hline
\end{tabular}

\end{table}

A dispersion relation can be approximated from equation~\eqref{eq:mix}:
\begin{equation}
\sin(\omega/\omega_{\rm p})\cos(\omega_{\rm g}/\omega)-q\sin(\omega_{\rm g}/\omega)\cos(\omega/\omega_{\rm p})=0
\label{eq:disper}
\end{equation}
where
\begin{equation}
\omega_{\rm g}=L\int_{r_{\rm a}}^{r_{\rm b}}\frac{N}{r}{\rm d}r\simeq\frac{2{\uppi}^2}{\Delta\Pi},~~~~\omega_{\rm p}=\left(\int_{r_{\rm c}}^R \frac{{\rm d}r}{c}\right)^{-1}\simeq2\Delta\nu.
\end{equation}
Although the dispersion relation differs from equation (25) in \cite{jcd12}, we follow derivations in \cite{jcd12} and rewrite equation~\eqref{eq:disper} as:
\begin{equation}
C(\omega) \cos (\omega_{\rm g} / \omega+ \Phi(\omega))=0,
\label{eq:newdisper}
\end{equation}
where $C(\omega)=\sqrt{\sin^2(\omega / \omega_{\rm p})+q^2 \cos^2(\omega/\omega_{\rm p})}$ and $\Phi(\omega)$ satisfies
\begin{align}
\label{eq:1cw}
 C(\omega) \cos\Phi(\omega) &= \sin(\omega / \omega_{\rm p}) \\
 C(\omega) \sin\Phi(\omega) &= q\cos(\omega / \omega_{\rm p}).
\end{align}
From equation~\eqref{eq:newdisper}, it is obvious that the eigenfrequencies satisfy
\begin{equation}
{\cal R}=\omega_{\rm g} / \omega + \Phi(\omega) = \left(n+\frac{1}{2}\right){\uppi}.
\end{equation}
Hence the frequency spacing between adjacent modes nearly satisfies 
\begin{equation}
\Delta \omega \simeq \Delta {\cal R} (\rm d {\cal R} / \rm d \omega)^{-1} = {\uppi} (\rm d {\cal R} / \rm d \omega)^{-1},
\end{equation}
and the corresponding period spacing is approximately given by
\begin{equation}
\begin{aligned}
\Delta \Pi & \simeq -\frac{2{\uppi} \Delta \omega}{\omega^2} = -\frac{2{\uppi}^2}{\omega^2} \left(\frac{\rm d {\cal R}} { \rm d \omega}\right)^{-1}\\
& = \frac{2{\uppi}^2}{\omega_{\rm g}} \left(1-\frac{\omega^2}{\omega_{\rm g}}\frac{\rm d \Phi}{\rm d \omega} \right)^{-1}.
\end{aligned}
\label{eq:1dp}
\end{equation}
The term outside the parentheses of equation~\eqref{eq:1dp} is the period spacing for pure g modes, and variations of $\Delta \Pi$ of mixed modes originate from the behaviour of $\Phi$ which is illustrated in \cite{jcd12}. $\Phi$ is almost constant and therefore its derivative is around zero except near acoustic resonances where $\Phi$ changes rapidly, causing a vigorous variation in the period spacing. At the centre of an acoustic resonance, the mode frequency is exactly $\nu_{n_{\rm p},\,l=1}$ in equation~\eqref{eq:mixnu}, and  $\Phi = {\uppi} / 2$, $\omega/\omega_{\rm p}=k{\uppi}$ for integer $k$. We introduce $\delta x = \omega/\omega_{\rm p} - k{\uppi}$ that represents how much a mode deviates from the centre of an acoustic resonance and expand $\Phi$ as $\delta \Phi = \Phi - {\uppi}/2$ in terms of $\delta x$. Equations (22) and (23) yield $\tan (\delta \Phi) = - q^{-1} \tan (\delta x)$, expanding which to the second order gives   
\begin{equation}
\delta \Phi \approx -\frac{\delta x}{\sqrt{\frac{2}{3}\delta x^2+q^2}},
\end{equation}
which leads to the derivative of $\Phi$ as
\begin{equation}
\begin{aligned}
\frac{\rm d \Phi}{\rm d \omega} &= \frac{\rm d }{\rm d \omega} (\delta \Phi + \frac{{\uppi}}{2}) \\
&=\frac{q^2}{\omega_{\rm p} (q^2+\frac{2}{3} \delta x^2)^{3/2}}
\end{aligned}
\label{eq:dphi}
\end{equation}
This leads to the final approximation to $\Delta \Pi$ after plugging equation~\eqref{eq:dphi} into equation~\eqref{eq:1dp}, 
\begin{equation}
\Delta\Pi \simeq \frac{2{\uppi}^2}{\omega_{\rm g}} \left(1+\frac{\omega^2}{\omega_{\rm g}\omega_{\rm p}}\frac{q^2} {(q^2+\frac{2}{3} \delta x^2)^{3/2}}\right)^{-1}.
\label{eq:dpfit}
\end{equation}
Therefore the variation of the period spacing is determined by the coupling strength and $\delta x$. The minimum period spacing occurs when the mode is extremely close to the centre of the acoustic resonance:
\begin{equation}
\Delta \Pi_{\rm min} \simeq \frac{2{\uppi}^2}{\omega_{\rm g}} \left(1+\frac{\omega^2}{q \omega_{g} \omega_{\rm p}}\right)^{-1}.
\label{eq:mindp}
\end{equation}

\begin{figure}
\resizebox{1.0\hsize}{!}{\includegraphics{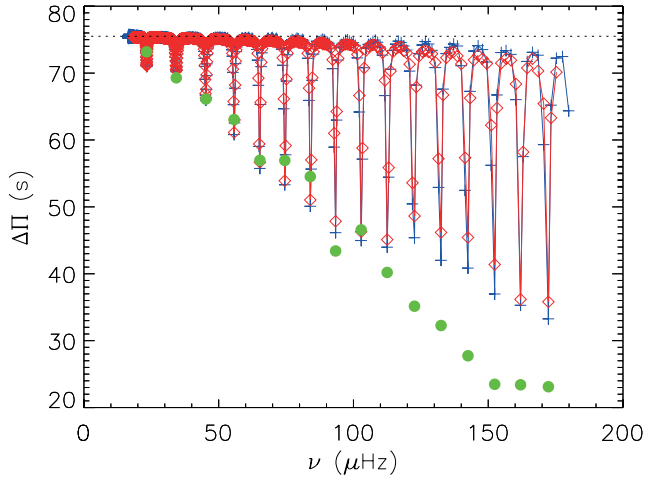}}
\caption{Same as the upper panel of Fig.~\ref{fg:dpfit} for model $M_{\rm 4}$.}
\label{fg:dpfita}
\end{figure}

In equation~\eqref{eq:dpfit}, $\omega_{\rm g}$ relates to the $\Delta \Pi$ of g-m modes, $\omega_{\rm p}$ corresponds to around twice $\Delta \nu$, and $\delta x \approx 0$ indicates a p-m mode. Therefore, equation~\eqref{eq:dpfit} clearly provides another approach to acquiring $\Delta \Pi$, $\Delta \nu$ as well as $q$ by fitting the period spacings if we have adequate g-m modes. To do this, we focused on $M_{\rm 3}$ which has a large amount of high-order g modes. Fig.~\ref{fg:dpfit} presents the fit to the period spacings of $M_{\rm 3}$. The period spacings of the g-m modes, which are located on the side of each dip, are very close to the theoretical $\Delta \Pi$ calculated from equation~\eqref{eq:dp}, while the p-m modes lie at the bottom of each dip, as a result of the increase in frequency of these modes. We fit each dip separately in the range [$\nu_{\rm p\hbox{-}m} - \Dnu / 2~;~\nu_{\rm p\hbox{-}m} + \Dnu / 2]$. Those modes at the bottom of each dip are regarded as p-m modes, which are mixed modes containing gravity mode character as well. They are also possible for detection if their $\zeta$ are large enough. As shown in Fig.~\ref{fg:dpfit}, the asymptotic dispersion relation leads to an excellent fit to the theoretical period spacings for low-frequency modes. However, when the g modes are not dense enough, $\omega_{\rm g}$ is underestimated and so therefore is $\Delta \Pi$. The results are displayed in Table~\ref{tb:dpfit}. The coupling strength relates to the depth and width of each dip. According to equation~\eqref{eq:mindp} the depth of each dip increases with decreasing $q$, but the width decreases reversely and therefore the chances of finding a g mode decrease too. However, in this exercise we found that the reductions of the period spacing are also very sensitive to the mode frequency. For instance, for low-frequency modes, the small value of the maximum reduction in period spacing of $M_{\rm 3}$ gives very high $q$ value, compared to the results from other approaches, though they lead to good period spacing calculated from $\omega_{\rm g}$ and mean frequency spacing derived from $\nu_{p\hbox{-}m}$ ($\Delta \Pi = 87.41 \pm 0.62$ s, $\Dnu = 18.04 \pm 0.16 ~\muHz$). We also analysed the even more evolved model $M_{\rm 4}$. The resulting $q$ are acceptable in this case thanks to sufficient g modes but the frequency dependence still exists (Table~\ref{tb:dpfita} and Fig.~\ref{fg:dpfita}). In summary, the outcome of the asymptotic dispersion relation works more robustly for more evolved stars where g modes are dense, but the results have a strong frequency dependence.

\begin{table}
\caption{Oscillation parameters for model $M_{\rm 4}$
obtained by fitting computed period spacings to equation (\ref{eq:dpfit}),
for individual acoustic resonances characterized by $\nu_{\rm p\hbox{-}m}$.
See the caption to Table \ref{tb:dpfit}.}
 \label{tb:dpfita}
 \begin{tabular}{@{}cccccccc}
 \hline
\small{$\nu_{p\hbox{-}m}$} &  \small{$\omega_{\rm g}$}  & \small{$\omega_{\rm p}$}  &  \small{$\Delta \Pi$}   &  \small{$\Dnu$}   &  \small{$q$}   &  \small{$q_{\rm int}$}   &  \small{$q_{\rm fit}$}  \\
 \small{($\muHz$)}   &  \small{($\rm s^{-1}$)}   & \small{($\muHz$)}   &  \small{(s)}  &  \small{($\muHz$)}  & & &  \\
 \hline
\small23.26    &  \small0.262   &  \small23.25   &  \small 75.38     &  \small 11.63   &  \small0.05   &  \small 0.005   & \small0.13\\
\small34.26    &  \small0.262    &  \small22.83   &  \small 75.24      &  \small 11.42   &  \small0.11   &  \small 0.012    &\small0.10\\
\small45.24    &  \small0.263   &  \small22.61    &  \small 75.14       &  \small 11.30   &  \small0.10    &  \small 0.021    &\small0.11\\ 
\small55.72    &  \small0.263   &  \small22.30   &  \small 75.01        &  \small 11.15   &  \small0.09    &  \small 0.031   &\small0.12\\
\small65.29  &  \small0.264   &  \small21.73     &  \small 74.88         &  \small 10.87    &  \small0.09    &  \small 0.041    & \small0.10\\
\small74.52   &  \small0.264   &  \small21.31   &  \small 74.76         &  \small 10.66    &  \small0.10    &  \small 0.051  & \small0.13\\
\small83.98    &  \small0.265    &  \small21.02    &  \small 74.59      &  \small 10.51     &  \small0.10    &  \small 0.060  & \small0.14\\
\small93.38    &  \small0.265    &  \small20.73    &  \small 74.35       &  \small 10.36     &  \small0.10   &  \small 0.069   & \small0.09\\
\small102.87   &  \small0.266    &  \small20.58   &  \small 74.27       &  \small 10.29     &  \small0.13    &  \small 0.078   & \small0.13\\
\small112.51   &  \small0.267     &  \small20.48   &  \small 73.99     &  \small 10.24    &  \small 0.13   &  \small 0.086   &\small0.11\\ 
\small122.64     &  \small 0.269   &  \small20.39    &  \small 73.48    &  \small 10.19      &  \small 0.13   &  \small 0.094   & \small0.10\\
\small132.51   &  \small0.269     &  \small20.34    &  \small 73.28      &  \small 10.17     &  \small 0.15    &  \small 0.102   & \small0.10\\
\small142.48     &  \small0.271   &  \small20.30   &  \small 72.92    &  \small 10.15     &  \small0.17      &  \small 0.109    & \small0.09\\ 
\small152.40    &  \small0.273     &  \small20.28    &  \small 72.40    &  \small 10.14     &  \small0.17     &  \small 0.116    & \small0.08\\
\small162.01   &  \small0.272      &  \small20.25    &  \small 72.49   &  \small 10.13      &  \small0.19     &  \small 0.124  & 0.09\\
\small172.33   &  \small0.279   &  \small20.24    &  \small 70.82    &  \small 10.12     &  \small0.16    &  \small 0.132   & \small0.10\\
\hline
\end{tabular}
\end{table}

\section{Conclusion}   

We have analysed the properties of the dipolar mixed modes ($l = 1$) of RG models. Mixed modes result from the coupling between the acoustic and gravity modes in evolved stars when the buoyancy frequency $N$ is high enough that frequencies of g modes are able to approach those of p modes. When a star evolves past the main sequence, the possible frequency upper limit that a g mode can get to increases dramatically because of the increase of the central density. In the beginning, oscillation modes are dominated by p modes and coupled with several g modes, and high-order pure g modes are only distributed over the low frequency region. Frequencies of p modes are bumped up to a slightly higher value when an avoided crossing happens. As the star ages, frequencies of more g modes grow to be as high as those of p modes. The dense g modes are then coupled with a few acoustic oscillation modes. The frequencies of g modes are diminished a little as a result of coupling in this case. The extent of the coupling can be measured by a coefficient called the coupling strength $q$ that corresponds to how close the g and p cavities are. The closer the two cavities are, the larger is $q$. 

The coupling actually affects every oscillation mode all the time in more evolved stars, at a level which also depends on the mode phases. \cite{mos12a} show an asymptotic relation of mixed modes derived from an implicit relation between the phases of coupled p and g modes. By using this asymptotic relation, we can measure the coupling strength and $\Delta \Pi$ to a very good degree for ascending-branch models by means of grid searching. In other words, the asymptotic relation is very helpful for the identification of mixed modes in observed power spectra. However, our fitting method failed to provide reasonable $q$ for early models where p modes still play the leading role. The fact that our grid searching only finds the best parameters without giving any uncertainties can be overcome with a more detail grid. 

It is well known that mixed modes have both p and g mode characters simultaneously, but at a different level. We find that for RG models when a mixed mode has considerable variation in frequency, the proportions of the contributions to its mode inertia from the outer region to the central core change significantly as well. Frequencies of those p-m modes with a large contribution ratio of mode inertia between the two regions ($\zeta \gg 1$) are very close to the pure p modes, which are obvious in the power spectrum. In their vicinities, several mixed modes with relatively smaller $\zeta$ ($\zeta \geqslant 1$) can be detected from observation, which have more g-cavity character and hence reveal more information of the central core. \cite{jcd12} estimated that the chance of finding such mixed modes is related to the width of the maximum reduction in period spacings, which is based on a study of the solutions of an asymptotic dispersion relation. Although we have a slightly different dispersion relation in this paper, an identical expression for the period spacing is obtained and has been employed to fit theoretical models, which gives excellent results for $\Delta \Pi$ and $\Dnu$. We confirm the prediction of \cite{jcd12} about the chance of finding mixed modes, but we also notice that the width and depth of the dips in period spacing depend on the combination of mode frequency and $q$. Our fitting to period spacing is able to supply uncertainties and also results in a good outcome of $q$ for more evolved stars, but unreasonably high value of $q$ for high-order g modes. The two fitting approaches can both provide robust outcomes for evolved stars when they are combined.
We note that, although our analysis is done for theoretical models, the method is also useful for real observations that provide adequate mixed modes. 

The value of $q$ inferred from the analysis of observed frequencies in principle
provides diagnostics of the stellar interior, supplementing $\Delta \nu$
and $\Delta \Pi$ which probe, respectively, the envelope and the core of 
the star.
This requires a better understanding of how $q$ is related to the structure 
of the star in the evanescent region.
The theoretical $q_{\rm int}$ from the integral expression
of equation~\eqref{eq:qori} is obviously inadequate for this purpose, given
the problems with the underlying asymptotic equation in the evanescent region.
A more careful analysis is required, taking into account the potential
singularities.
A second limitation in the analysis is the use of the Cowling approximation,
neglecting the perturbation to the gravitational potential. 
We have tested the importance of this by determining the coupling strength
computed by fitting frequencies computed in the Cowling approximation,
comparing with results based on the full set of equations, and showing 
a significant reduction in $q$, by about 0.02.
We note that the Cowling approximation 
could be avoided in the present case of dipolar modes by using the
exact second-order equation developed by \citet{Takata2005}.
This definitely deserves further investigation.

\section{Acknowledgements}
We are very grateful to MarieJo Goupil for a conversation which provided
crucial inspiration for the present analysis and we thank the referee for pertinent comments on an earlier version of the manuscript which greatly improved the presentation.
We also thank Beno\^{i}t Mosser for providing observed frequencies and helpful discussions.
Funding for the Stellar Astrophysics Centre is provided by the Danish National Research Foundation (grant agreement no.: DNRF106). The research is supported by the ASTERISK project (ASTERoseismic Investigations with SONG and \kepler\ ) funded by the European Research Council (grant agreement no.: 267864).

\end{document}